# Classification of Visual Perception and Imagery based EEG Signals Using Convolutional Neural Networks

Ji-Seon Bang
*Dept. Brain and Cognitive Engineering*
*Korea University*
Seoul, Republic of Korea
js_bang@korea.ac.kr

Ji-Hoon Jeong
*Dept. Brain and Cognitive Engineering*
*Korea University*
Seoul, Republic of Korea
jh_jeong@korea.ac.kr

Dong-Ok Won
*Dept. Artificial Intelligence Convergence*
*Hallym University*
Chuncheon, Republic of Korea
dongok.won@hallym.ac.kr

*Abstract*—Recently, visual perception (VP) and visual imagery (VI) paradigms are investigated in several brain-computer interface (BCI) studies. VP and VI are defined as a changing of brain signals when perceiving and memorizing visual information, respectively. These paradigms could be alternatives to the previous visual-based paradigms which have limitations such as fatigue and low information transfer rates. In this study, we analyzed VP and VI to investigate the possibility to control BCI. First, we conducted a time-frequency analysis with event-related spectral perturbation. In addition, two types of decoding accuracies were obtained with convolutional neural network to verify whether the brain signals can be distinguished from each class in the VP and whether they can be differentiated with VP and VI paradigms. As a result, the 6-class classification performance in VP was 32.56% ($\pm$7.07) and the binary classification performance which classifies two paradigms was 90.16% ($\pm$9.69).

## I. INTRODUCTION

Brain-computer interface (BCI) allows users to control external devices using their intentions, which are decoded from brain signals [1]–[4]. BCI involves several paradigms such as motor imagery (MI) [5], [6], event-related potential (ERP) [7], steady-state visually evoked potential (SSVEP) [8], [9], and rapid serial visual presentation (RSVP) [10], [11]. Also, there have been conducted several visual-based studies [12], [13]. Recently, visual imagery (VI) and visual perception (VP) are investigated as novel paradigms in several BCI studies [14], [15]. VP is defined as a changing of brain signal when someone perceives certain visual information and VI is defined as a fluctuation of brain signal when someone memory certain visual information [16].

There have been several electroencephalography (EEG) studies based on VI and VP. Kosmyna *et al.* [14] analyzed what extent can the VI and VP be differentiated. They performed VP and VI while paying attention and imagining two images, a flower and a hammer. The following binary classification performances were presented: VP vs VI, VP of two classes, resting-state vs VP, and resting-state vs VI. In a study by Sousa *et al.* [15], VP and VI of 3-class were proposed. They provided static dot and moving dot as stimulation conditions. In specific, a static dot and a moving dot in the vertical axis, and a moving dot in vertical and horizontal axes are the 3-class that are used for the experiment. They conducted the time frequency analysis for both VP and VI using event-related spectral perturbation (ERSP) to represent the brain responses in the whole brain area. They found a decrease of alpha-band (8-13 Hz) power, especially in the area from the parietal to occipital during VP. Also, they found an increase of alpha-band power, especially in the area from the frontal to central during VI paradigm.

Besides, VP and VI-based magnetoencephalography study also exists. In a study by Dijkstra *et al.* [17], the author compared VP and VI to provide insights for the understanding of the neural mechanisms of visual processing. They showed that VI decoding becomes significant later compared to VP. This finding suggests that the entire visual representation is activated at once during VI. Also, they found out that whereas VP was characterized by high temporal specificity, VI showed low temporal specificity. A study by Spyropoulos *et al.* [13] represented the manifestation of visual tasks with electrocorticography. The author showed that a theta band (3-8 Hz) is predominant in visual attention tasks from the experiment with two macaque monkeys. They found this result by recording local field potential from the visual cortex, V1, V2, V4, and temporal-occipital (TEO) region.

Meanwhile, in the process of decoding the brain signal,

20xx IEEE. Personal use of this material is permitted. Permission from IEEE must be obtained for all other uses, in any current or future media, including reprinting/republishing this material for advertising or promotional purposes, creating new collective works, for resale or redistribution to servers or lists, or reuse of any copyrighted component of this work in other works.

This work was partly supported by Institute of Information & Communications Technology Planning & Evaluation (IITP) grant funded by the Korea government (MSIT) (No. 2015-0-00185, Development of Intelligent Pattern Recognition Softwares for Ambulatory Brain Computer Interface, No. 2017-0-00451, Development of BCI based Brain and Cognitive Computing Technology for Recognizing User's Intentions using Deep Learning, No. 2019-0-00079, Artificial Intelligence Graduate School Program(Korea University)).

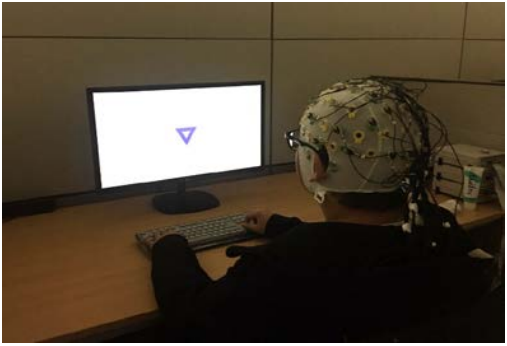

Fig. 1. The experimental environment of VP and VI paradigm.

the feature extraction methods are particularly important, as there exist large inter- and intra-subject variability in the EEG. Although classical feature extraction methods such as common spatial pattern (CSP) [18] and filter bank common spatial pattern (FBCSP) [19] are exist, the convolutional neural network (CNN) showed enhanced performance in the BCI researches recently [20]–[26]. For example, Cecotti *et al.* [20], first adopted CNN for decoding of P300, Stober *et al.* [21], classified music rhythm, and Manor *et al.* [22], classified RSVP with CNN. Sakhavi *et al.* [23], Schirrmeister *et al.* [24], and Bang *et al.* [25] adopted CNN in MI-based BCIs and Kwak *et al.* [26], adopted CNN in SSVEP based BCIs.

In this study, we investigate the possibility to control BCI. For this purpose, we conducted VP and VI experiments and analyzed them with 4 subjects. Six different colored shapes were provided for the visual stimuli. With the obtained brain signal, we conducted a time-frequency analysis. Also, from the temporal-spatial input feature of the signal, we obtained two types of decoding accuracy with CNN. First was a 6-class classification in the VP paradigm to verify whether the brain signal can be differentiated with each class. The second was classification which classifies VP and VI paradigms to verify whether the brain signal can be differentiated with each paradigm.

## II. MATERIALS AND METHODS

### A. Participants

4 healthy participants (3 males and 1 female; aged 24 to 29) participated in the experiment. All participants were informed about the paradigm before the experiment. The experimental protocols were reviewed and approved by the Institutional Review Board (IRB) at Korea University [KUIRB- 2020-0056-01].

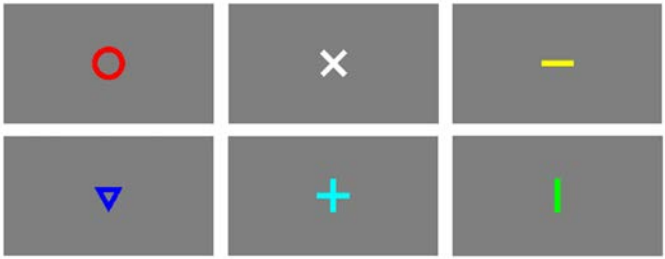

Fig. 2. Six visual stimuli which are provided for VP and VI.

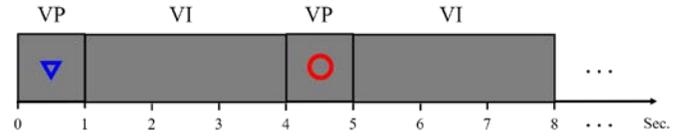

Fig. 3. The experimental paradigm.

### B. Visual Stimuli

Figure 2 shows six stimuli used in this experiment. The six colored shapes were provided as visual stimuli for the experiment refer to Shen *et al.* [27]; red circle, white cross, yellow horizontal line, blue triangle, cyan plus, and green vertical line. The visual stimuli were presented at the center of the monitor. The size of the visual stimuli was 8cm × 8cm and the distance between subject and monitor was 80cm so that the degrees of viewing angle for the visual stimuli were 3°. This angle is matching with a foveal vision which refers to the center of the field of vision. When the image is focused in this area, the visual acuity is at its highest. By setting the viewing angle as 3°, the shape would be located to the foveal vision, so that the subject can recognize visual stimuli accurately without moving their eyes.

### C. Experimental Paradigm

Figure 3 represents the experimental paradigm. For the VP experiment, subjects were instructed to pay attention to the shape which is appeared on the monitor for 1,000 ms. After the shape disappears, the monitor becomes blank, and subjects were instructed to imagine the shape for 3,000 ms which was appeared before. This process is VI experiment. Six visual stimuli appear in random order in one cycle. When all six visual stimuli appear and disappear, one cycle is completed.

One cycle is defined as one trial and it lasts for 24 sec. The whole experiment is composed of 150 trials so that the total experiment time is one hour. To keep subjects pay attention, the experiment setup was designed to receive keyboard input from the subjects to move on to the next trial.

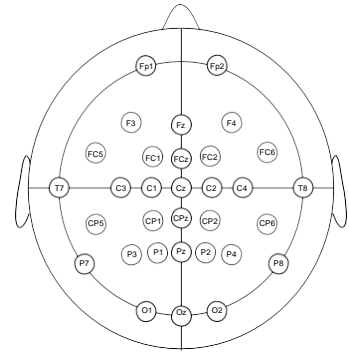

Fig. 4. The electrode channel montage which is used in the experiment.

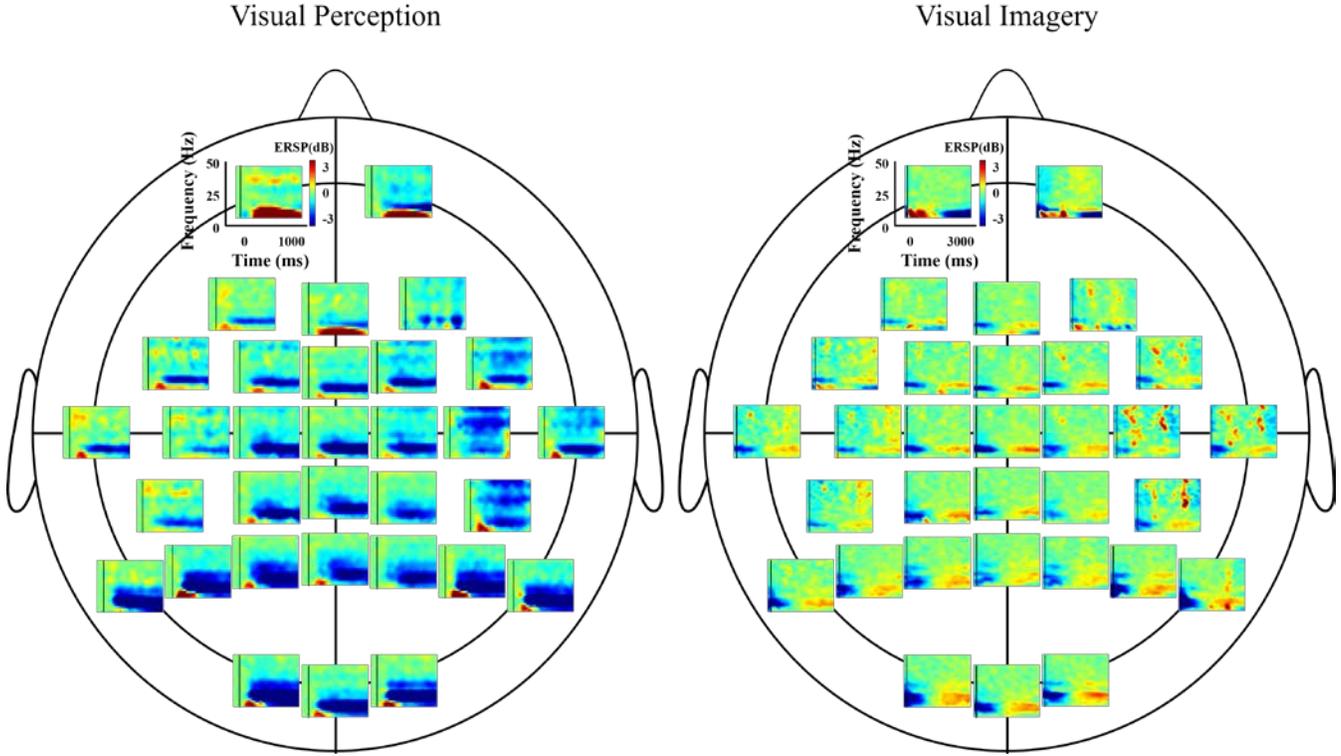

Fig. 5. Brain responses while conducting 1,000 ms VP (left) and 3,000 ms VI (right). The figure was represented based on ERSP.

*D. EEG Recording*

EEG data were recorded with 32 Ag/AgCl electrodes channels using a 1,000 Hz sampling rate according to the 10-20 international system (Fig. 4). EEG amplifier (BrainAmp, Brain Products) were used for the experiment. Reference and ground electrodes were fixed at posterior auricular and nasion, respectively.

*E. Time-Frequency Analysis*

We conducted a time-frequency analysis with ERSP with the OpenBMI toolbox (http://openbmi.org) [28]. The EEG data were band-pass filtered between 0.5 and 50 Hz for all channels, across entire trials. To analyze VP, the filtered data were segmented from -100 to 1,000 ms, based on when the visual stimuli appeared. For VI, the signal from -100 to 3,000 ms was segmented based on when the visual stimuli disappeared. From each spectral estimate, the mean baseline log power spectrum was subtracted, to visualize power changes across the frequency range. This process produced baseline-normalized ERSP [29].

*F. Preprocessing*

We filtered and segmented both signals which were obtained while performing VP and VI for the classification. The signal from the VP was filtered with theta band (3-8 Hz) refer to [13] and was segmented from 0 to 1,000 ms based on the time the visual stimulus appeared. The signal obtained from the VI was filtered with the same band with VP and was segmented from 1,000 to 2,000 ms based on the time the visual stimulus was disappeared. The reason why we used not 0 but 1,000 for the start point was to remove VP effects.

To classify six classes in the VP paradigm, we used the preprocessed signal from VP. In addition, to classify each paradigm, we used both the preprocessed signal from VP and VI. With the segmented signal, we constructed temporal-spatial input feature for CNN so that the size of the feature would be time points (1,000) × channels (32).

*G. Data Classification with CNN*

The CNN network consists of 2 convolutional layers, 1 exponential linear unit (ELU) layer [30], 1 average-pooling layer, and 1 fully-connected layer. We designed the network so that each convolutional layer to be applied to the temporal features and the spatial features respectively. The first convolutional layer is adopted across the temporal domain, with 25 × 1 sized filter. The second convolutional layer followed by ELU layer is adopted across spatial domain, with 1 × 32 sized filter. The filter size of average-pooling layer was 75 × 1 and the stride size was 15 × 1. This structure is the same for both 6-class classification in VP and binary classification which classifies VP and VI, except for the number of neurons in the classification layer (six and two, respectively). We employed Adam-optimizer [31] to optimize the cost. The learning rate was set to 0.001 and the iteration was set to 100. We obtained

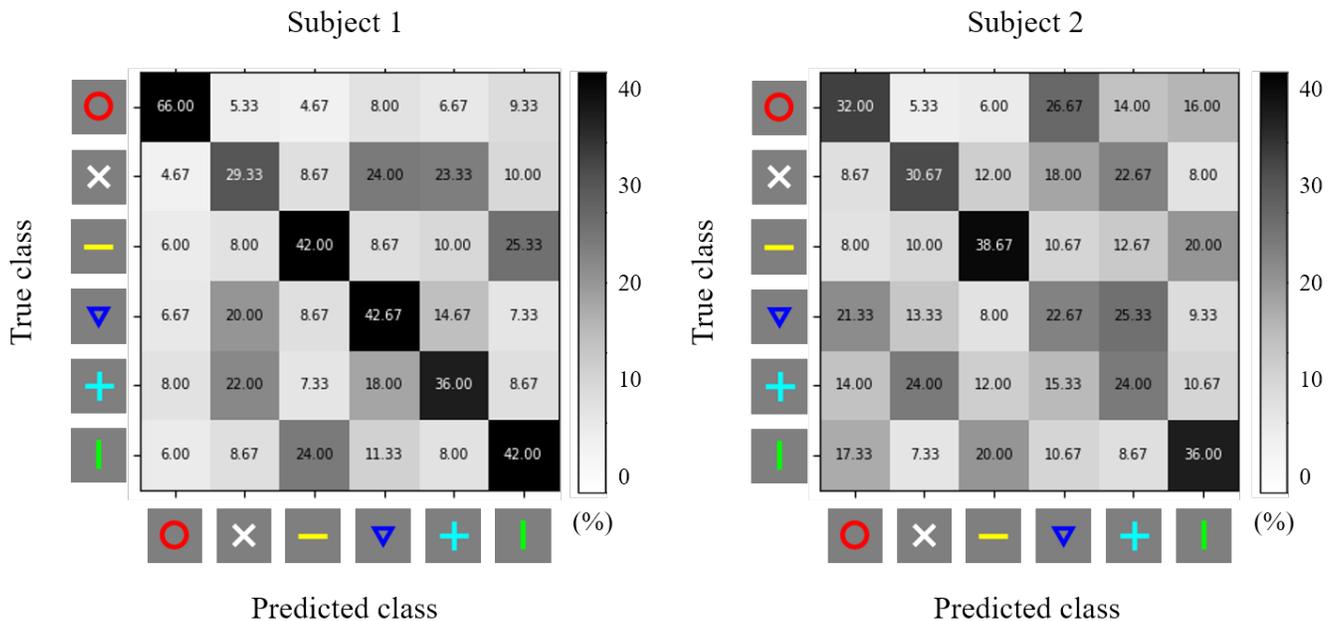

Fig. 6. The confusion matrix of VP for the subject 1 (left) and the subject 2 (right). Classification accuracy with 10-fold cross-validation is used.

the decoding accuracy by performing 10-fold cross-validation for both classification results.

## III. RESULTS

### A. ERSP Analysis

We conducted a time-frequency analysis with ERSP. The analysis was conducted with all participants. The grand average of the brain responses while conducting VP (Figure 5 (left)) shows a significant decrease of theta and alpha-band power in the occipital area where the visual cortex is located. The analysis was also conducted for VI. The grand average of the brain responses while conducting VI (Figure 5 (right)) shows a decrease of alpha-band power until around 1,000 ms and the increase of alpha-band power after 1,000 ms in the whole-brain area.

### B. Classification Results

In Figure 6, the confusion matrixes of 6-class visual perception for the subject 1 (left) and the subject 2 (right) are represented. Subject 1 and 2 are chosen as they achieved the highest and the next highest performance, respectively. The result was obtained based on 10-fold cross-validation.

TABLE I
ACCURACY, PRECISION, RECALL, AND F1 SCORE FROM THE
CLASSIFICATION RESULT BETWEEN VP AND VI

|          | Sub1  | Sub2  | Sub3  | Sub4  | mean  |
|----------|-------|-------|-------|-------|-------|
| Accuracy | 0.988 | 0.961 | 0.772 | 0.884 | 0.902 |
| Precision| 0.990 | 0.963 | 0.762 | 0.890 | 0.901 |
| Recall   | 0.987 | 0.958 | 0.790 | 0.878 | 0.903 |
| F1 score | 0.988 | 0.960 | 0.776 | 0.884 | 0.902 |

The figure shows that the subject 1 classified the first class (red circle) as the highest performance with 66.00%, and the subject 2 classified the third class (yellow horizontal line) as the highest performance with 38.67%. On the other hand, the subject 1 classified the second class (white cross) as the lowest performance with 29.33%, and the subject 2 classified the fourth class (blue triangle) as the lowest performance with 22.67%.

In Figure 7, the average performance of the 6-class classification in VP and the average performance of the binary classification which classifies VP and VI are presented. The chance level of two types of classification is also presented with the red line, which is 16.66% and 50%, respectively. The results show that the decoding accuracies are 32.56% and 90.16%, respectively.

In Table I, the classification result which classifies VP and VI are shown, based on the accuracy, precision, recall, and F1 score. The definition of each index is as follows:

$$Accuracy = \frac{(TP + TN)}{(TP + FN + FP + TN)} \quad (1)$$

$$Precision = \frac{TP}{(TP + FP)} \quad (2)$$

$$Recall = \frac{TP}{(TP + FN)} \quad (3)$$

$$F1\ score = 2 \times \frac{(Precision \times Recall)}{(Precision + Recall)} \quad (4)$$

where TP denotes true positive rate, TN denotes true negative rate, FN denotes false negative rate, and FP denotes false positive rate.

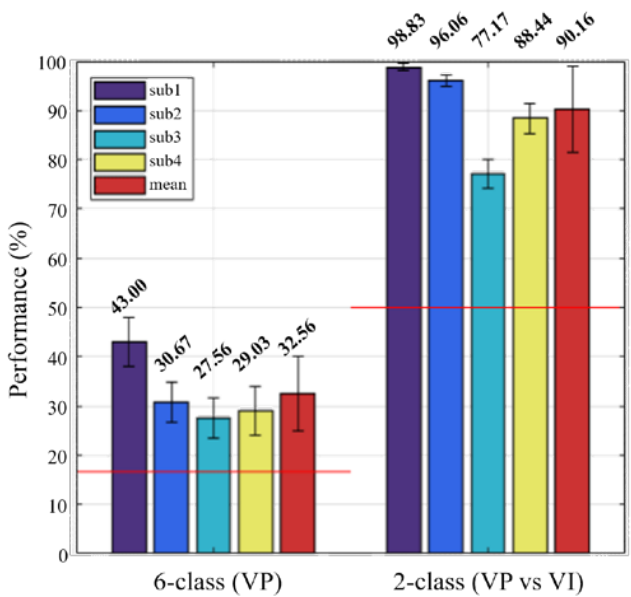

Fig. 7. The binary classification performance which classifies VP and VI (right) and 6-class classification performance in VP (left).

The result was obtained with 10-fold cross-validation. In the result, subject 1 achieved the highest values for each index with 0.988, 0.990, 0.987, and 0.988, respectively. Subject 3 showed the lowest values for each index with 0.772, 0.762, 0.790, and 0.776, respectively.

## IV. Discussion

In Figure 5 (left), the brain responses of the VP are represented. From the result, we found a significant decrease in theta and alpha-band power in the occipital area. These patterns are similar to the results which are represented from the previous studies [15]. It is well known that visual stimuli induce a decrease in the alpha-band power in the occipital area. This result indicates that the brain signal while performing VP is related to the perception of visual stimuli.

In Figure 5 (right), the brain responses of the VI are represented. From the result, we found that the decrease of alpha-band power followed by the increase of alpha-band power. The increase of alpha-band power is similar to that from the previous studies [15]. Whereas, for the decrease of alpha-band power which is appeared at the beginning, it appears that the signal generated by the VP is affected. This is because the VI experiment is right behind the VP experiment.

To avoid interference from VP, updating the experimental paradigm would be needed for the further improvement. To ensure that the VI paradigm does not come directly behind the VP paradigm, putting a random noise image between VP and VI can be one solution. We can expect the random noise image would minimize the impact of VP so that it will improve the quality of brain signals while performing VI.

In Figure 6, meanwhile, there is a difference between classes that are well classified and not well classified depending on each subject. The results show that the classes with the highest performance and the lowest performance were different for each subject. This indicates that to achieve stable performance, a feature extraction model should extract subject-dependent feature representation related to the visual character. Thus, further research should be to find subject-dependent visual features for each subject.

## V. Conclusion

In this study, we conducted experiments of VP and VI with six different colored plane figures. From the experimental output, we analyzed brain responses during VP and VI. From the result, we found a significant decrease of theta and alpha-band power while conducting VP and found that the increase of alpha-band power follows the decrease of alpha-band power while conducting VI. These results are similar to the results from previous work so that we could conclude that the brain signal from both paradigms is related to the perception and imagining of visual stimuli, respectively. Also, we obtained the decoding accuracy of 6-class classification in the VP paradigm, and 2-class classification which classifies VP and VI paradigms with deep learning methods. The results were 32.56% and 90.16%, respectively. These results indicate that each class of VP can be classified and the brain signal while conducting VP and VI can be also differentiated as they showed significant performance, especially for the binary classification. From the result, we can conclude that the use of the proposed method is beneficial for the classification and prediction of the brain states from visual tasks (i.e. VP and VI) when developing comfortable BCI systems. Further study would be to control BCI systems through VP and VI or analyzing them in other paradigms such as magnetic resonance imaging (MRI) [32], positron emission tomography (PET) [33], or multi-modal paradigm [34].